%% file: main.tex
\RequirePackage{amsmath} 
\RequirePackage{accents}
\RequirePackage{amssymb}
\documentclass[english]{eptcs}
\usepackage[T1]{fontenc}
\usepackage[utf8]{inputenc}
\usepackage{underscore}           
\usepackage{amsthm}
\usepackage{babel}
\usepackage{xargs}
\usepackage{stmaryrd}
\usepackage{paralist}
\usepackage{booktabs}
\usepackage{enumitem}
\usepackage{pdfpages}
\usepackage{subcaption}
\usepackage{algorithm}
\usepackage{algorithmic}
\newcommand{\SWITCH}[1]{\STATE \textbf{switch} (#1)}
\newcommand{\ENDSWITCH}{\STATE \textbf{end switch}}
\newcommand{\CASE}[1]{\STATE \textbf{case} #1\textbf{:} \begin{ALC@g}}
	\newcommand{\ENDCASE}{\end{ALC@g}}

\newcommand{\DEFAULT}{\STATE \textbf{default:} \begin{ALC@g}}
	\newcommand{\ENDDEFAULT}{\end{ALC@g}}
\newcommand{\DEFAULTLINE}[1]{\STATE \textbf{default:} }
\usepackage{stackengine}
\usepackage{xifthen}
\usepackage{color}
\definecolor{dkgreen}{rgb}{0,0.6,0}
\definecolor{gray}{rgb}{0.5,0.5,0.5}
\definecolor{mauve}{rgb}{0.58,0,0.82}
\usepackage{tcolorbox}
\tcbuselibrary{skins,breakable,theorems}
\newtcolorbox{code}{boxrule=0.5pt,left=0mm,right=0mm,top=0mm,bottom=0mm}
\usepackage{mathtools}
\usepackage{wasysym}
\tcbuselibrary{skins,breakable}
\usepackage[mode=image]{standalone}
\usepackage{fixme}
\fxsetup{
	final,
	author=,
	theme=color
}
\usepackage{relsize}
\usepackage{factum}
\usepackage{tikz}
\newtcolorbox{mymathbox}[2][]
{ams align*, enhanced, bottom=0mm, top=0mm, boxrule=0.5pt, colback=white, coltitle=black, title=#2, attach boxed title to top left={xshift=-2mm,yshift=-4mm}, boxed title style={size=small,frame hidden,colback=black!10}, #1}
\usetikzlibrary{matrix,shapes,decorations.pathreplacing,calc,patterns,positioning,decorations.pathmorphing,automata,fit}
\usepackage{listings}
\lstdefinelanguage{AspectJ}[]{Java}{
	morekeywords={declare, pointcut, aspect, before, around, after, returning, throwing, call, execution, this, target, args, within, withincode, get, set, initialization, preinitialization, staticinitialization, handler, adviceexecution, cflow, cflowbelow, if, proceed},
	moredelim=[is][\textcolor{darkgray}]{\%\%}{\%\%},
	moredelim=[il][\textcolor{darkgray}]{§§}
}
\input{macros/math}

\input{macros/model}

\title{Detecting Architectural Erosion \\ using Runtime Verification}
\author{Diego Marmsoler\textsuperscript{\url{https://orcid.org/0000-0003-2859-7673}}
	\institute{Technische Universit\"at M\"unchen, Germany}
	\email{diego.marmsoler@tum.de}
	\and
	Ana Petrovska\textsuperscript{\url{https://orcid.org/0000-0001-6280-2461}}
	\institute{Technische Universit\"at M\"unchen, Germany}
	\email{ana.petrovska@tum.de}
}
\def\titlerunning{Detecting Architectural Erosion using RV}
\def\authorrunning{D. Marmsoler and A. Petrovska}
\PassOptionsToPackage{hyphens}{url}
\usepackage{hyperref}
\include{macros/templates}
\begin{document}
	\maketitle
	\input{abstract}
	\input{introduction}
	\input{background}
	\input{rtverification}
	\input{approach}
	\input{evaluation}	
	\input{rw}
	\input{conclusion}
	\newpage
	\bibliographystyle{eptcs}
	\bibliography{references}
\end{document}

%% file: macros/math.tex
\usepackage{accents}

\newcommand{\inlineeqno}[1]{\refstepcounter{equation}\label{#1}(\theequation)}

\newcommand{\defeq}[1][0]{\overset{\text{def}}{\ifthenelse{#1=0}{\quad}{}=\ifthenelse{#1=0}{\quad}{}}}
\newcommand{\defiff}{\overset{\text{def}}{\iff}}

\newcommand{\pset}[1]{\wp{\left(#1\right)}}

\newcommand{\vdir}[2]{\ensuremath{[#1\mapsto #2]}}

\newcommand{\sinfin}[1]{\ensuremath{(#1)^\infty}}

\theoremstyle{definition}
\newtheorem{definition}{Definition}[section]
\newtheorem{example}[definition]{Example}
\theoremstyle{plain}

%% file: macros/model.tex
\newcommand{\port}{\mathcal{P}}
\newcommand{\mess}{\mathcal{M}}
\newcommand{\ptype}{\mathcal{T}}
\newcommand{\pVal}[1]{\ensuremath{\overline{#1}}}

\newcommand{\iface}[1][]{\ensuremath{(\ifIP#1,\ifOP#1)}}
\newcommand{\ifIP}{\ensuremath{\mathit{I}}}
\newcommand{\ifOP}{\ensuremath{\mathit{O}}}
\newcommand{\ifAllSymb}{\ensuremath{\mathit{IF}}}
\newcommand{\ifAll}[1]{\ensuremath{\ifAllSymb_{#1}}}
\newcommand{\ifIn}[1]{\ensuremath{\mathsf{in}(#1)}}
\newcommand{\ifOut}[1]{\ensuremath{\mathsf{out}(#1)}}
\newcommand{\ifPort}[1]{\ensuremath{\mathsf{port}(#1)}}

\newcommand{\cmpAll}{\ensuremath{\mathcal{C}}}
\newcommand{\cmpIn}[1]{\ensuremath{\mathsf{in}(#1)}}
\newcommand{\cmpOut}[1]{\ensuremath{\mathsf{out}(#1)}}

\newcommand{\cmpPort}[1]{\ensuremath{\mathsf{port}(#1)}}

\newcommand{\configuration}[1][]{\ensuremath{(\cnfComp#1,\cnfConn#1,\cnfVal#1)}}
\newcommand{\cnfComp}{\ensuremath{C'}}
\newcommand{\cnfConn}{\ensuremath{N}}
\newcommand{\cnfVal}{\ensuremath{\mu}}
\newcommand{\cactive}[2]{\ensuremath{\lvert#2\rvert_{#1}}}
\newcommand{\asactive}[1]{\ensuremath{\mathit{CMP}_{#1}}}
\newcommand{\asconn}[1]{\ensuremath{\mathit{CN}_{#1}}}
\newcommand{\asval}[1]{\ensuremath{\mathit{val}_{#1}}}
\newcommand{\cnfcmp}[2]{\ensuremath{\mathsf{cmp}_{#1}^{#2}}}
\newcommand{\archAllSymb}{\ensuremath{\mathit{AS}}}
\newcommand{\archAll}[2]{\ensuremath{\archAllSymb_{#1}^{#2}}}

\newcommand{\allTraces}[2]{\ensuremath{\sinfin{\archAll{#1}{#2}}}}



%% file: macros/templates.tex
\newlength{\tmpwidth}
\newlength{\cpos}

\newenvironmentx{dtstemp}[3][1=250pt, 3]{
	\newboolean{var}
	\setlength{\tmpwidth}{#1}
	\setlength{\cpos}{15pt}

	\newcommand{\var}[2]{%
		\addtolength{\cpos}{10pt}
		\node[anchor=base west] at (0,-\cpos) {\ifthenelse{\boolean{var}}{}{\textbf{flex}}};
		\node[anchor=base west] at (0.8cm,-\cpos) {##1$\colon$};
		\node[anchor=base east, align=right] at (\tmpwidth,-\cpos) {##2};		
		\addtolength{\cpos}{5pt}
		\setboolean{var}{true}		
	}
	
	\newcommand{\axiom}[2][]{%
		\addtolength{\cpos}{10pt}		
		\node [anchor=base west] at (0,-\cpos) {##2};
		\ifthenelse{\equal{##1}{}}{}{
			\node [anchor=base east, text width=1.5cm, align=right] at (\tmpwidth,-\cpos) {\inlineeqno{##1}};}
		\addtolength{\cpos}{5pt}		
	}

	\begin{tikzpicture}
	\draw [thick] (0,0) -- (\tmpwidth,0);
	\node [anchor=base west] at (0,-10pt) {\textbf{DTSpec} #2};
	\ifthenelse{\equal{#3}{}}{}{%
	\node [align=right,anchor=base east, text width=0.5*\tmpwidth] at (\tmpwidth,-10pt) {\textbf{imports} #3}};
	\draw [thin, double, double distance=2pt] (0,-15pt) -- (\tmpwidth,-15pt);
}
{
	\draw[thick] (0,-\cpos) -- (\tmpwidth,-\cpos);
	\end{tikzpicture}		
}

\newenvironmentx{pstemp}[3][1=4cm,3]{
	\setlength{\tmpwidth}{#1}
	\setlength{\cpos}{0pt}

	\begin{tikzpicture}	
	\draw [thick] (0,\cpos) -- (\tmpwidth,\cpos);
	\addtolength{\cpos}{10pt}	
	\node [anchor=base west] at (0,-\cpos) {\textbf{PSpec} #2};
	\node [anchor=base east, align=right] at (\tmpwidth,-\cpos) {\ifthenelse{\equal{#3}{}}{}{\textbf{imports} #3}};
	\addtolength{\cpos}{5pt}
	\draw [thin, double,double distance=2pt] (0,-\cpos) -- (\tmpwidth,-\cpos);
}
{
	\draw[thick] (0,-\cpos) -- (\tmpwidth,-\cpos);
	\end{tikzpicture}		
}

\newenvironmentx{bhvspec}[4][1=200pt]{
	\setlength{\tmpwidth}{#1}
	\setlength{\cpos}{0pt}	
	\def\func{30}
	\def\bfirst{0}
	\def\rfirst{0}	
	
	\newcommand{\bhvvar}[2]{%
		\addtolength{\cpos}{5pt}
		\node [anchor=base west] at (0,-\cpos) {\ifthenelse{\bfirst=0}{\textbf{flex}}{}};
		\node [anchor=base west] at (1cm,-\cpos) {##1$\colon$};
		\node [anchor=base east,align=right] at (\tmpwidth,-\cpos) {##2};
		\edef\bfirst{1}	
		\addtolength{\cpos}{5pt}
	}
	\newcommand{\bhvrig}[2]{%
		\addtolength{\cpos}{5pt}
		\node [anchor=base west] at (0,-\cpos) {\ifthenelse{\rfirst=0}{\textbf{var}}{}};
		\node [anchor=base west] at (1cm,-\cpos) {##1$\colon$};
		\node [anchor=base east,align=right] at (\tmpwidth,-\cpos) {##2};
		\edef\rfirst{1}	
		\addtolength{\cpos}{5pt}
	}
	\newcommand{\bhvaxiom}[2][]{%
		\addtolength{\cpos}{\heightof{##2}}
		\addtolength{\cpos}{2pt}		
		\node [anchor=base west] at (0,-\cpos) {##2};
		\ifthenelse{\equal{##1}{}}{}{
			\node [anchor=base east, text width=1.5cm,align=right] at (\tmpwidth,-\cpos) {\inlineeqno{##1}};}
		\addtolength{\cpos}{2pt}
		\addtolength{\cpos}{\depthof{##2}}		
	}

	\begin{tikzpicture}	
	\draw [thick] (0,\cpos) -- (\tmpwidth,\cpos);
	\addtolength{\cpos}{10pt}
	\node [anchor=base west] at (0,-\cpos) {\textbf{BSpec} #2};
	\node [align=right,anchor=base east, text width=0.5*\tmpwidth] at (\tmpwidth,-10pt) {\textbf{for} #3 \textbf{of} #4};
	\addtolength{\cpos}{5pt}	
	\draw [thin, double,double distance=2pt] (0,-\cpos) -- (\tmpwidth,-\cpos);
	\addtolength{\cpos}{5pt}	
}
{
	\draw [thick] (0,-\cpos) -- (\tmpwidth,-\cpos);
	\end{tikzpicture}
}

\newenvironmentx{ctspec}[3][1=200pt]{
	\setlength{\tmpwidth}{#1}
	\setlength{\cpos}{0pt}	
	\def\func{30}
	\def\bfirst{0}
	\def\rfirst{0}
	
	\newcommand{\ctvar}[2]{%
		\addtolength{\cpos}{5pt}
		\node [anchor=base west] at (0,-\cpos) {\ifthenelse{\bfirst=0}{\textbf{flex}}{}};
		\node [anchor=base west] at (1cm,-\cpos) {##1$\colon$};
		\node [anchor=base east,align=right] at (\tmpwidth,-\cpos) {##2};
		\edef\bfirst{1}	
		\addtolength{\cpos}{5pt}
	}
	\newcommand{\ctrig}[2]{%
		\addtolength{\cpos}{5pt}
		\node [anchor=base west] at (0,-\cpos) {\ifthenelse{\rfirst=0}{\textbf{var}}{}};
		\node [anchor=base west] at (1cm,-\cpos) {##1$\colon$};
		\node [anchor=base east,align=right] at (\tmpwidth,-\cpos) {##2};
		\edef\rfirst{1}	
		\addtolength{\cpos}{5pt}
	}

	\newcommand{\ctline}[1][solid]{%
		\draw[##1] (0,-\cpos) -- (\tmpwidth,-\cpos);
	}
	\newcommand{\ctaxiom}[2][]{%
		\addtolength{\cpos}{\heightof{##2}}
		\addtolength{\cpos}{2pt}		
		\node [anchor=base west] at (0,-\cpos) {##2};
		\ifthenelse{\equal{##1}{}}{}{
			\node [anchor=base east, text width=1.5cm,align=right] at (\tmpwidth,-\cpos) {\inlineeqno{##1}};}
		\addtolength{\cpos}{2pt}
		\addtolength{\cpos}{\depthof{##2}}
	}

	\begin{tikzpicture}	
	\draw [thick] (0,\cpos) -- (\tmpwidth,\cpos);
	\addtolength{\cpos}{10pt}
	\node [anchor=base west] at (0,-\cpos) {\textbf{ASpec} #2};
	\node [align=right,anchor=base east, text width=0.5*\tmpwidth] at (\tmpwidth,-10pt) {\textbf{for} #3};
	\addtolength{\cpos}{5pt}	
	\draw [thin, double,double distance=2pt] (0,-\cpos) -- (\tmpwidth,-\cpos);
	\addtolength{\cpos}{5pt}	
}
{
	\draw [thick] (0,-\cpos) -- (\tmpwidth,-\cpos);
	\end{tikzpicture}
}

%% file: abstract.tex
\begin{abstract}
	The architecture of a system captures important design decisions for the system.
    Over time, changes in a system's implementation may lead to violations of specific design decisions.
    This problem is common in industry and known as architectural erosion.
    Since it may have severe consequences on the quality of a system, research has focused on the development of tools and techniques to address the presented problem.
    As of today, most of the approaches to detect architectural erosion employ static analysis techniques.
    While these techniques are well-suited for the analysis of static architectures, they reach their limit when it comes to dynamic architectures.
    Thus, in this paper, we propose an alternative approach based on runtime verification.
    To this end, we propose a systematic way to translate a formal specification of architectural constraints to monitors, which can be used to detect violations of these constraints.
    The approach is implemented in Eclipse/EMF, demonstrated through a running example, and evaluated using two case studies.\looseness-1
\end{abstract}

%% file: introduction.tex
\section{Introduction}
A system's architecture captures major design decisions made about the system to address its requirements.
However, changes in the implementation may sometimes change the original architecture, and some of the design decisions might become invalid over time.
This situation, sometimes called architectural erosion~\cite{Garlan1995}, may have severe consequences on the quality of a system and is a common, widespread problem in industry~\cite{Eick2001,Godfrey2000}.
Thus, research has proposed approaches and tools to detect architectural erosion which, as of today, focus mainly on the analysis of static architectures and therefore employ static analysis techniques~\cite{Coverity2019,Deissenboeck2010,Klocwork2019,Lattix2019,Sonargraph2019,structure1012019}.\looseness-1

However, recent trends in computing, such as mobile and ubiquitous computing, require architectures to adapt dynamically:
new components may join or leave the network and connections between them may change over time.
Thereby, architectural changes can happen at runtime and depend on the state of the components.
Thus, analysis of erosion for dynamic architectures requires the analysis of component behavior, and therefore it is only difficult, not to say impossible, to detect with static analysis techniques.

Consider, for example, the following scenario:
An architect, to satisfy important memory requirements, decides to implement the Singleton pattern to restrict the number of active components of a specific type.
Since the developers are not familiar with this memory requirement, they modify the code in a way that they create multiple instances of the corresponding type.
Detecting the corresponding architectural violation with static analysis tools is difficult, not to say impossible.

Thus, traditional approaches to detect architectural erosion may reach their limits when it comes to dynamic architectures.
To address this problem, we propose a novel approach to detect architectural erosion, based on runtime verification~\cite{Lavery2017}:\looseness-1
\begin{compactitem}
    \item First, architectural assertions are formally specified in \textsc{FACTum}~\cite{Marmsoler2018c}, a language for the formal specification of dynamic architectural constraints.
    \item Second, code instrumentation and monitors are generated from the specification.
    \item Finally, the monitors are used to detect architectural violations at runtime.
\end{compactitem}
To this end, we developed two algorithms to map a given \textsc{FACTum} specification to corresponding events and LTL-formul\ae{} over these events.

We evaluated the approach on two case studies from different domains.
In the first case study, we applied the approach in a controlled environment for the analysis of an open-source Java application in the domain of Business Information Systems.
Accordingly, we implemented the algorithms for Java applications, specified architectural constraints in \textsc{FACTum}, and generated monitors and code instrumentation.
Finally, we executed the system and observed it for violations.

The second case study was executed in a real, industrial setting, in which we applied the approach for the analysis of a proprietary C application in the automotive domain.
To this end, we first implemented the algorithms for C applications.
Again, we specified architectural constraints in \textsc{FACTum} and generated corresponding monitors and code instrumentation.
Lastly, we executed the system observing it for architectural violations.

With this paper, we provide two major contributions:
\begin{compactitem}
    \item We describe a systematic way to detect architectural erosion for dynamic architectures using runtime verification techniques and demonstrate its applicability using a running example.
    To this end, we also describe two algorithms for mapping a \textsc{FACTum} specification to corresponding events and LTL-formul\ae{} over these events.
    \item We show the feasibility of using runtime verification to detect architectural erosion through two case studies.
\end{compactitem} 
The paper is structured as follows:
We first provide some background on \textsc{FACTum} (Sect.~\ref{sec:factum}) and runtime verification (Sect.~\ref{sec:rv}).
Then, we introduce our approach and demonstrate each step using a simple, running example (Sect.~\ref{sec:approach}).
In (Sect.~\ref{sec:evaluation}) we present the two case studies.
Finally, we discuss related work (Sect.~\ref{sec:rw}) and conclude the paper with a summary and limitations that lead to potential future work (Sect.~\ref{sec:conc}).

%% file: background.tex
\section{Specifying Dynamic Architectures in \textsc{FACTum}}\label{sec:factum}
\textsc{FACTum}~\cite{Marmsoler2018c} is an approach for the formal specification of constraints for dynamic architectures.
It consists of a formal system model for dynamic architectures and techniques to specify constraints over this model.
\textsc{FACTum} is also implemented in terms of an Eclipse/EMF application called \textsc{FACTum} Studio~\cite{Marmsoler2018f}, which supports a user in the development of specifications.

\subsection{System Model}
In our model~\cite{Marmsoler2016a,Marmsoler2016}, components communicate with each other by exchanging messages over ports.
Thus, we assume the existence of set $\mess$, containing all \emph{messages}, and set $\port$, containing all \emph{ports}, respectively.
Moreover, we postulate the existence of a type function
\begin{equation}\label{eq:type}
\ptype\colon \port \to \pset{\mess}
\end{equation}
which assigns a set of messages to each port.

\textit{Port valuations.}
Ports are means to exchange messages between a component and its environment.
This is achieved through the notion of port valuation.
Roughly speaking, a valuation for a set of ports is an assignment of messages to each port.
\begin{definition}[Port valuation]\label{def:pval}
	For a set of ports $P\subseteq \port$, we denote with $\pVal{P}$ the set of all possible, type-compatible \emph{port valuation}, formally:
	\begin{equation*}
	\pVal{P}\defeq \Big\{\mu\in \big(P\to\pset{\mess}\big) \mid \forall p\in P\colon \mu(p)\subseteq\ptype(p) \Big\}\enspace.
	\end{equation*}
	Moreover, we denote by $\vdir{p_1,p_2,\hdots}{M_1,M_2,\hdots}$ the valuation of ports $p_1, p_2, \hdots$ with sets $M_1$, $M_2$, $\hdots$, respectively. For singleton sets we shall sometimes omit the set parentheses and simply write $\vdir{p_1, p_2, \hdots}{m_1,m_2,\hdots}$\enspace.
\end{definition}
In our model, ports may be valuated by \emph{sets} of messages, meaning that a component can send/receive a set of messages via each of its ports at each point in time.
A component may also send no message at all, in which case the corresponding port is valuated by the empty set.\looseness-1

\textit{Interfaces.}
The ports which a component may use to send and receive messages are grouped into so-called interfaces.
\begin{sloppypar}
	\begin{definition}[Interface]\label{def:iface}
		An \emph{interface} is a pair $\iface$, consisting of \emph{disjoint} sets of \emph{input ports} $\ifIP\subseteq\port$ and \emph{output ports} $\ifOP\subseteq\port$.
		The set of all interfaces is denoted by $\ifAll{\port}$.
		For an interface $\mathit{if}=\iface$, we denote by
		\begin{compactitem}
			\item $\ifIn{\mathit{if}}\defeq \ifIP$ the set of input ports,
			\item $\ifOut{\mathit{if}}\defeq \ifOP$ the set of output ports, and
			\item $\ifPort{\mathit{if}}\defeq \ifIP \cup \ifOP$ the set of all interface ports.
		\end{compactitem}	
	\end{definition}
\end{sloppypar}


\textit{Components.}\label{sec:model:components}
For the purpose of this paper, we assume the existence of a set of components $(\cmpAll_\mathit{if})_{\mathit{if}\in\ifAll{\port}}$.
A \emph{component port} is a port combined with the corresponding component identifier.
Thus, for a family of components $(\cmpAll_\mathit{ct})_{\mathit{if}\in\ifAll{\port}}$ over a set of interfaces $\ifAll{\port}$, we denote by:
\begin{compactitem}
	\item $\cmpIn{\cmpAll}\defeq\bigcup_{c\in \cmpAll}\left(\{c\}\times \cmpIn{\mathit{c}}\right)$, the set of \emph{component input ports},
	\item $\cmpOut{\cmpAll}\defeq\bigcup_{c\in \cmpAll}\left(\{c\}\times \cmpOut{\mathit{c}}\right)$, the set of \emph{component output ports},	
	\item $\cmpPort{\cmpAll}\defeq\cmpIn{\cmpAll}\cup\cmpOut{\cmpAll}$, the set of all \emph{component ports}.
\end{compactitem}

Moreover, we may lift the typing function (introduced for ports in Eq.~\ref{eq:type}), to corresponding component ports: \[\ptype((c,p))\defeq\ptype(p)\enspace.\]
Finally, we can generalize our notion of port valuation (Def.~\ref{def:pval}) for \emph{component ports} $\mathit{CP}\subseteq \cmpAll\times\port$ to a so-called \emph{component port valuation}:
\begin{equation*}
\pVal{\mathit{CP}}\defeq \Big\{\mu\in \big(\mathit{CP}\to\pset{\mess}\big) \mid \forall \mathit{cp}\in \mathit{CP}\colon \mu(\mathit{cp})\subseteq\ptype(\mathit{cp}) \Big\}\enspace.
\end{equation*}
To better distinguish between ports and component ports, in the following, we shall use $p$, $q$, $\mathit{pi}$, $\mathit{po}$, \dots; to denote ports and $\mathit{cp}$, $\mathit{cq}$, $\mathit{ci}$, $\mathit{co}$, \dots; to denote component ports.

\subsubsection{Architecture Snapshots.}
In our model, an architecture snapshot \emph{connects} ports of \emph{active} components.
\begin{definition}[Architecture snapshot]\label{def:aconf}
	An \emph{architecture snapshot} is a triple $\configuration$, consisting of:
	\begin{compactitem}
		\item a set of active components $\cnfComp\subseteq \cmpAll$,
		\item a connection $\cnfConn\colon \cmpIn{\cnfComp}\to\pset{\cmpOut{\cnfComp}}$, such that types of connected ports are compatible:
		\begin{equation}\label{eq:type:consistency}
		\forall \mathit{ci}\in\cmpIn{\cnfComp}\colon\bigcup_{\mathit{co}\in \cnfConn(\mathit{ci})} \ptype(\mathit{co})\subseteq\ptype(\mathit{ci})\enspace,\textrm{ and}
		\end{equation}
		\item a component port valuation $\cnfVal\in \pVal{\cmpPort{\cnfComp}}$\enspace.
	\end{compactitem}
	We require connected ports to be consistent in their valuation, i.e., if a component provides messages at its output port, these messages are transferred to the corresponding, connected input ports:
	\begin{equation}\label{eq:aconf:conn}
	\forall \mathit{ci}\in \cmpIn{\cnfComp} \colon \cnfConn(\mathit{ci})\neq\emptyset\implies \cnfVal(\mathit{ci})=\bigcup_{\mathit{co}\in \cnfConn(\mathit{ci})} \cnfVal(\mathit{co})\enspace.
	\end{equation}
	Note that Eq.~\eqref{eq:type:consistency} guarantees that Eq.~\eqref{eq:aconf:conn} does not violate type restrictions.
	The set of all possible architecture snapshots is denoted by $\archAll{\ptype}{\cmpAll}$.
	
	For an architecture snapshot $\mathit{as}=\configuration\in \archAll{\ptype}{\cmpAll}$, we denote by
	\begin{compactitem}
		\item $\asactive{\mathit{as}}\defeq \cnfComp$ the set of active components and with $\cactive{\mathit{as}}{c}\defiff c\in\cnfComp$, that a component $c\in \cmpAll$ is \emph{active} in $\mathit{as}$,
		\item $\asconn{\mathit{as}}\defeq \cnfConn$, its connection, and
		\item $\asval{\mathit{as}}\defeq \cnfVal$, the port valuation.
	\end{compactitem}
	Moreover, given a component $c\in \cnfComp$, we denote by
	\begin{equation}
	\cnfcmp{\mathit{as}}{c}\in \pVal{\ifPort{c}} \defeq \Big(\lambda p\in \ifPort{c}\colon \cnfVal\big((c,p)\big)\Big)
	\end{equation}
	the valuation of the component's ports.
\end{definition}
\noindent
Note that $\cnfcmp{\mathit{as}}{c}$ is well-defined iff $\cactive{\mathit{as}}{c}$.

Moreover, note that connection $\cnfConn$ is modeled as a set-valued function from component input ports to component output ports, meaning that:
\begin{compactenum}
	\item input/output ports can be connected to several output/input ports, respectively, and
	\item not every input/output port needs to be connected to an output/input port (in which case the connection returns the empty set).
\end{compactenum}

Moreover, note that by Eq.~\eqref{eq:aconf:conn}, the valuation of an input port connected to many output ports is defined to be the \emph{union} of all the valuations of the corresponding, connected output ports.

\begin{sloppypar}
\begin{example}[Architecture snapshot]\label{ex:config}
		Figure~\ref{fig:config} shows a conceptual representation of an architecture snapshot $\configuration$, consisting of:
	\begin{compactitem}
		\item active components $\cnfComp=\{c_1,c_2,c_3\}$ with corresponding interfaces;
		\item connection $\cnfConn$, defined as follows:
		\begin{itemize}
			\item $\cnfConn((c_2,i_1))=\{(c_1,o_1)\}$,
			\item $\cnfConn((c_3,i_1))=\{(c_1,o_2)\}$,
			\item $\cnfConn((c_2,i_2))=\{(c_3,o_1)\}$, and
			\item $\cnfConn((c_1,i_0))=\cnfConn((c_2,i_0))=\cnfConn((c_3,i_0))=\emptyset$; and
		\end{itemize}
		\item component port valuation $\vdir{(c_1,o_0),(c_2,i_1),(c_3,o_1),\cdots}{\mathtt{M_3},\mathtt{M_5},\mathtt{M_3},\cdots}$.
	\end{compactitem}
\end{example}
\end{sloppypar}			
\begin{figure}
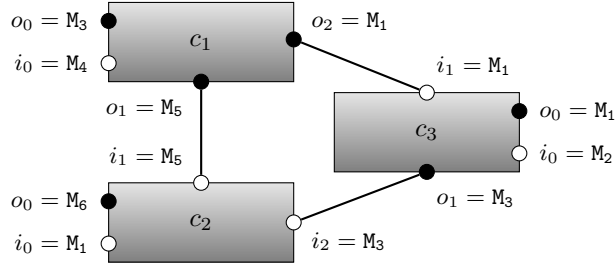

	\centering
	\includestandalone{img/config}
	\caption{Architecture snapshot consisting of components $c_1$, $c_2$, and $c_3$; a connection between ports $(c_2,i_1)$ and $(c_1,o_1)$, $(c_2,i_2)$ and $(c_3,o_1)$, and $(c_3,i_1)$ and $(c_1,o_2)$; and valuations of component ports.}
	\label{fig:config}
\end{figure}

\subsubsection{Architecture Traces.}
An \emph{architecture trace} consists of a series of snapshots of an architecture during system execution. Thus, an architecture trace is modeled as a stream of architecture snapshots at certain points in time.

\begin{sloppypar}
\begin{definition}[Architecture trace]\label{def:atrace}
	An \emph{architecture trace} is an infinite stream $t\in\allTraces{\ptype}{\cmpAll}$.
\end{definition}
\end{sloppypar}
\begin{example}[Architecture trace]\label{ex:ctrace}
	Figure~\ref{fig:trace} shows an architecture trace $t \in \allTraces{\ptype}{\cmpAll}$ with corresponding architecture snapshots $t(0)=k_0$, $t(1)=k_1$, and $t(2)=k_2$. architecture snapshot $k_0$, for example, is described in Ex.~\ref{ex:config}.
\end{example}
\begin{figure*}
	\centering
	\includestandalone[width=\textwidth]{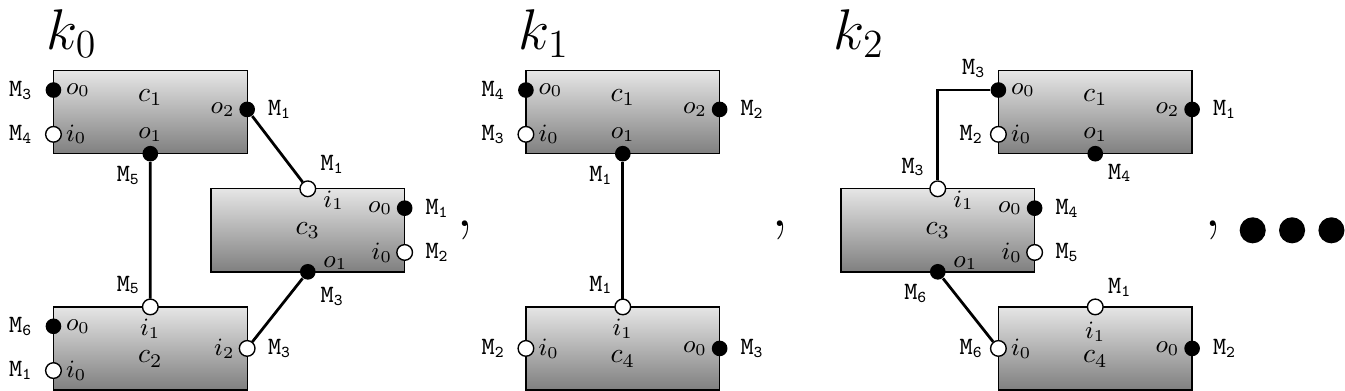}
	\caption{The first three architecture snapshots of an architecture trace.}
	\label{fig:trace}
\end{figure*}	

\subsection{Specifying Constraints for Dynamic Architectures}
\begin{sloppypar}
	\textsc{FACTum} provides several techniques to support the formal specification of constraints for dynamic architectures:
	\begin{compactitem}
		\item First, the data types involved in an architecture are specified in terms of algebraic specifications~\cite{Broy1996,Wirsing1990}.
		\item Then, a set of interfaces is specified graphically using architecture diagrams.
		\item Finally, a set of architectural assertions is added to specify constraints about component activation and deactivation as well as interconnection.
	\end{compactitem}
\end{sloppypar}
\noindent
A \textsc{FACTum} specification comes with a formal semantics in a denotational style.
To this end, each specification is interpreted by a corresponding set of architecture traces (as introduced in Def.~\ref{def:atrace}).

\textit{Architecture diagrams.}
Architecture diagrams~\cite{Marmsoler2019} are a graphical formalism to specify interfaces.
To this end, interfaces are represented by rectangles with their ports denoted by empty (input) and filled (output) circles.
An example of an architecture diagram can be found in Fig.~\ref{fig:diag}.

\textit{Specifying architectural constraints.}
Architectural constraints are specified in terms of \emph{architecture trace assertions}~\cite{Marmsoler2019}.
These are a type of first order linear temporal logic formul\ae{}~\cite{pnueli1977temporal}, with variables denoting components and some special terms and predicates:
\begin{compactitem}
	\item With $\caVal{c}{p}$, we denote that port $p$ of a component $c$ is currently active.
	\item With $\caPVal{c}{p}$, for example, we denote the valuation of port $p$ of a component $c$.
	\item With $\caActive {c}$ we denote that component $c$ is currently active.
	\item With $\caConn{c}{o}{c'}{i}$ we denote that output port $o$ of $c$ is connected to input port $i$ of $c'$.
\end{compactitem}
An example of an architecture trace assertion can be found in Fig.~\ref{fig:bc:act}.
A formal semantics is provided by Marmsoler and Gidey in~\cite{Marmsoler2019a}.


%% file: rtverification.tex
\section{Runtime Verification}\label{sec:rv}
Runtime verification (RV) is tightly related to and has its origins in model checking~\cite{clarke1999model,emerson2008beginning}. RV is a dynamic analysis method aiming at checking whether a run of the system under scrutiny satisfies a given correctness property~\cite{leucker2009brief}. RV deals with observed executions as the system generates them. Consequently, it applies to black box systems for which no system model is available, or to systems where the system model changes during the execution. 

In RV the correctness of a system is usually checked by a monitor. Therefore, through the literature, runtime verification is also referred to as runtime monitoring. 
``A monitor is a device that reads a finite trace and yields a certain verdict''~\cite{leucker2009brief}. In runtime verification, monitors are generated automatically from high-level specifications, and they need to be designed in a way that they consider system's executions incrementally. The specifications are usually formulated with temporal logic, for example, linear temporal logic (LTL) \cite{bauer2006model,pnueli1977temporal}. In the simplest form, a monitor decides if a program execution satisfies a particular correctness property or not. The system under analysis, as well as the generated monitor, are executed simultaneously~\cite{bauer2006monitoring}. Namely, the monitor observes the system's behavior. If the monitor detects that property is violated, then it returns a corresponding alarm signal. RV considers only the detection of violations of the correctness properties of a system. Even though RV does not necessarily affect the execution of a program, monitoring allows remedial action to be taken upon the detection of incorrect or faulty behavior.
In RV, often it is distinguished between online and offline methods; in online, a data stream is directly fed into the monitor, whereas in offline monitoring, data is provided from a log file.
In this paper, to generate monitors in the first and the second case study we used JavaMOP and LTL3 respectively. In the following subsections, we will briefly explain both of the tools. 

\textit{JavaMOP.}
Monitoring-Oriented Programming (MOP)~\cite{chen2007mop}, is a formal software development and analysis framework for RV. In MOP the developer specifies desired properties, or generates monitors, using specification formalisms. The monitors are integrated with the user-defined code into the original system, and the code is executed whenever the properties are violated or validated at runtime. This allows the original system to check its dynamic behaviors during execution and it reduces the gap between formal specification and implementation by allowing them to form a system together. Once a violation is detected, user-defined actions are triggered.
JavaMOP is MOP-based analysis and runtime verification system for Java, using AspectJ~\cite{Foundation2019} for instrumentation. Expressive requirements specification formalisms can be included in the framework via logic plug-ins, allowing not only to refer to the current state but also to both past and future states~\cite{chen2005java,Runtimeverification2019}.

\textit{LTL3 tools.}
$LTL_{3}$~\cite{bauer2006monitoring,bauer2011runtime}, is a 3-valued linear time temporal logic that can be interpreted over finite traces based on the standard semantics of LTL for infinite traces. $LTL_{3}$ shares the syntax with LTL but deviates in its semantics for finite traces.  The readings of the finite traces and the creation of 3-valued LTL semantics can be automated, and accordingly directly deployed as a runtime verification. 
$LTL_{3}$ Tools are a collection of programs to generate Finite State Machine through LTL formula. $LTL_{3}$ Tools takes an LTL formula and outputs a 3-valued corresponding monitor~\cite{Bauer2019}.

%% file: approach.tex
\section{Approach}\label{sec:approach}

Figure~\ref{fig:approach} depicts our approach for the verification of dynamic architecture constraints:
\begin{figure}
	\centering
	\includestandalone[width = .7\textwidth]{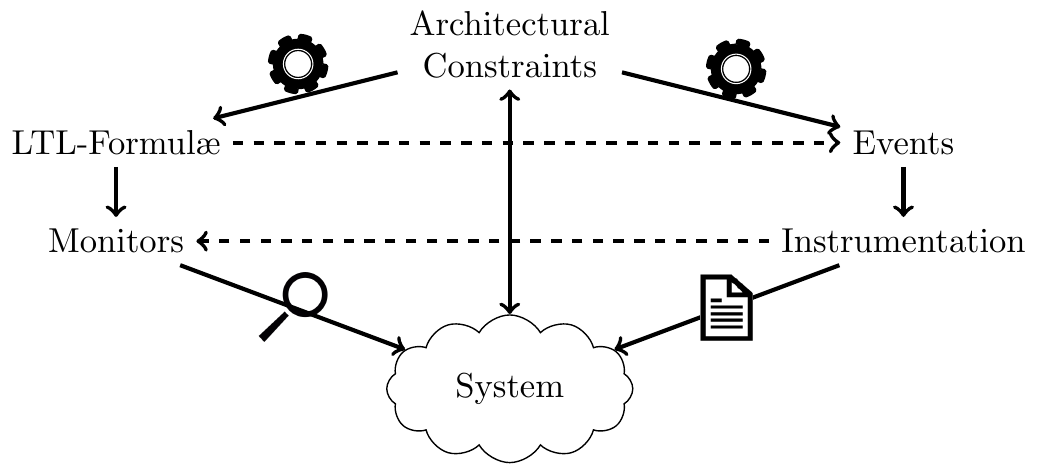}
	\caption{Runtime Verification of Architectural Constraints\label{fig:approach}}
\end{figure}
As a first step, a set of architectural constraints is specified in \textsc{FACTum}, consisting of a specification of component types $\mathit{CT}$ and a specification of architectural constraints $\mathit{AS}$ (see Sect.~\ref{sec:factum}).
The specification is then used to create two types of artifacts:
a set of events that will be monitored and a set of LTL-formul\ae{} based on these events.
The events are then used to create corresponding instrumentation code for the system to notify the monitor about the occurrence of events.
The LTL-fromul\ae{} are used to generate corresponding monitors to supervise the system for violations of the constraints.
While the first steps are mostly independent of the target platform of the system under test, the latter steps depend on the concrete platform of the system under test.

\subsection{Running Example: Online Shop}\label{sec:example}
To demonstrate the approach, we use a running example from the domain of business information systems.
In the following, we depict an excerpt of a possible implementation of such a shop in an object-oriented programming language.
In this paper, we are only interested in two classes: \textit{Baskets} and \textit{Items}.
The following listing sketches the implementation of class \textit{Basket}:
\begin{lstlisting}
public class Basket {
	private List items;
	public void addItem(String name, Integer price) {
		Item it = new Item();
		it.setName(name);
		it.setPrice(price);
		items.add(it);
	}    
}
\end{lstlisting}
The basket contains a collection of items and a method to add items to the list by using their name and price.

Thereby, the class \textit{Item} is implemented as follows:
\begin{lstlisting}
public class Item {
	private String name;
	private Integer price;
	
	public void setName(String nm) {
		this.name = nm;
	}

	public void setPrice(String pr) {
		this.price = pr;
	}
}
\end{lstlisting}
Each item has a name and a price and methods to modify them.

\subsection{Specifying the Property}
To analyze a system for architectural violations, we must first formulate corresponding architectural assertions.
For our webshop, one possible assertion we would like to check, could be as follows:
\begin{tcolorbox}\centering
	Whenever a user adds an item to its basket, a corresponding component of type \textit{Item} is created and initialized with the correct price and name.
\end{tcolorbox}

\begin{figure}
	\centering
	\includestandalone[width=.6\textwidth]{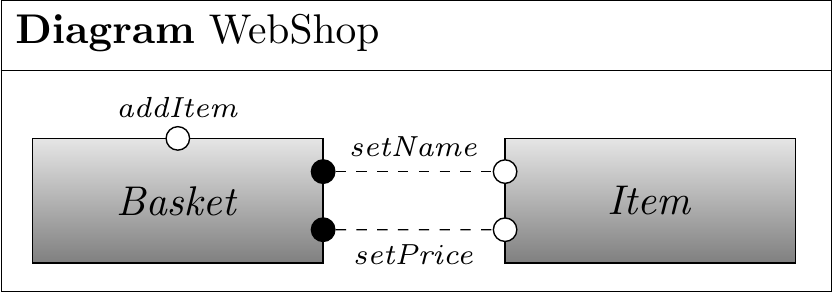}
	\caption{Architecture diagram for WebShop}\label{fig:diag}
\end{figure}

To formalize the property in \textsc{FACTum}, we first need to create a corresponding architecture diagram.
Figure~\ref{fig:diag} depicts the architecture diagram for our webshop example.
It depicts the two types of components required to specify the property: a \textit{Basket} and an \textit{Item}.
The \textit{Basket} has one input port \textit{addItem}, which can be used to trigger the addition of a new item.
Moreover, it has two output ports \textit{setName} and \textit{setPrice} to set the name and price of an item.
The \textit{Item}, on the other hand, has two input ports \textit{setName} and \textit{setPrice} to set a name and price, respectively.

The architecture diagram depicts the interfaces for our architecture.
The architecture constraint described above can now be formalized over these interfaces in terms of architectural assertions (Sect.~\ref{sec:factum}).
A possible formalization of the constraint is depicted in Fig.~\ref{fig:bc:act}.
Roughly speaking, the specification requires that, whenever a component $\mathit{bs}$ of type $\mathit{Basket}$ receives a message $(n,p)$ on its input port $\mathit{addItem}$ (Eq.~\eqref{eq:ws:act1}), a component of type $\mathit{Item}$ is created (Eq.~\eqref{eq:ws:act2}) and initialized with price $\mathit{p}$ (Eq.~\eqref{eq:ws:act3}) and name $\mathit{n}$ (Eq.~\eqref{eq:ws:act4}).
To prevent potential security issues, the connection constraints provided in Eq.~\eqref{eq:ws:act3} and Eq.~\eqref{eq:ws:act4}, require that the initialization is indeed done through the basket component itself.
Note that the specification imposes an ordering for the initialization of an item: first, the price has to be set and then the name.
\begin{figure}
	\centering
	\begin{ctspec}[340pt]{AddItem}{WebShop}
		\ctrig{$\mathit{n}$}{$\mathit{String}$}
		\ctrig{$\mathit{p}$}{$\mathit{Integer}$}
		\ctrig{$\mathit{it}$}{$\mathit{Item}$}
		\ctrig{$\mathit{bs}$}{$\mathit{Basket}$}		
		\ctline[dashed]
		\ctaxiom[eq:ws:act1]{$\ctglobally\Big(\caPVal{\mathit{bs}}{\mathit{addItem}}=(n,p)\ctimplies$}
		\ctaxiom[eq:ws:act2]{$\quad\ctnext\big(\caActive{\mathit{it}}\big)$}
		\ctaxiom[eq:ws:act3]{$\quad\caland~\ctnext\ctnext\big(\caPVal{\mathit{bs}}{\mathit{setPrice}}=p\caland\caConn{\mathit{bs}}{\mathit{setPrice}}{\mathit{it}}{\mathit{setPrice}}\big)\Big)$}
		\ctaxiom[eq:ws:act4]{$\quad\caland~\ctnext\ctnext\ctnext\big(\caPVal{\mathit{bs}}{\mathit{setName}}=n\caland\caConn{\mathit{bs}}{\mathit{setName}}{\mathit{it}}{\mathit{setName}}\big)$}	
	\end{ctspec}
	\caption{Architectural Constraints for Web-shop.\label{fig:bc:act}}
\end{figure}%

\subsection{Generating Events to Monitor}
We can now use the specification of component types to create a set of events which we would like to monitor.
Algorithm~\ref{alg:inst} shows a systematic way to do so:
For each type of component, corresponding activation events are created (Line 2).
Further, each input port results in the creation of corresponding execution events (Lines 4-8).
Output ports, on the other hand, result in call events (Lines 9-12).
Finally, each pair of input and output port (of the same name) results in the creation of a corresponding call event (Lines 13-21).
\renewcommand{\algorithmicrequire}{\textbf{Input:}}%
\renewcommand{\algorithmicensure}{\textbf{Output:}}%

\begin{algorithm}[t]
	\caption{Mapping \textsc{FACTum} to events for instrumentation}\label{alg:inst}
	\begin{algorithmic}[1]
		\REQUIRE a set $\mathit{CT}$ of component types
		\ENSURE a set of events
		\FORALL {$\mathit{ct}\in \mathit{CT}$}
		\STATE {create \emph{activation} event $\mathit{ct}\_\mathtt{activation}(\mathtt{ct})$} \COMMENT {where $\mathtt{ct}$ is a variable of type $\mathit{ct}$}
		\FORALL {ports $p$ of $\mathit{ct}$}
		\IF {$p$ is an input port}
		\STATE {create \emph{execution} event $\mathit{ct}\_p\_\mathtt{execution}(\mathtt{ct})$}
		\STATE {create \emph{execution} event $\mathit{ct}\_p\_\mathtt{execution}(\mathtt{ct}, \mathtt{params})$}\STATE\COMMENT {where $\mathtt{params}$ is a list of variables corresponding to the type of $p$}
		\ENDIF
		\IF {$p$ is an output port}
		\STATE {create \emph{call} event $\mathit{ct}\_p\_\mathtt{call}(\mathtt{ct})$}
		\STATE {create \emph{call} event $\mathit{ct}\_p\_\mathtt{call}(\mathtt{ct}, \mathtt{params})$}
		\ENDIF
		\FORALL {$\mathit{ct'}\in \mathit{CT}$}
		\IF {$\mathit{ct'}\neq\mathit{ct}$}
		\FORALL {ports $p'$ of $\mathit{ct'}$}				
		\IF {$p=p'$ and $p$ is an input port}
		\STATE {create \emph{call} event $\mathit{ct}\_\mathtt{call}\_\mathit{ct'}\_p(\mathtt{ct}, \mathtt{ct'})$}
		\ENDIF
		\ENDFOR
		\ENDIF											
		\ENDFOR
		\ENDFOR
		\ENDFOR
	\end{algorithmic}
\end{algorithm}

Instrumenting our web-shop example would require to monitor $14$ types of events derived from the specification of component types depicted in Fig.~\ref{fig:diag}:
\begin{inparaenum}[(i)]
	\item two types of activation events for $\mathit{Basket}$ components and $\mathit{Item}$ components,
	\item $10$ types of execution and call events (with and without parameters) for $\mathit{addItem}$, $\mathit{setName}$, and $\mathit{setPrice}$, and
	\item two types of connection events for $\mathit{setName}$ and $\mathit{setPrice}$.
\end{inparaenum}
The concrete types of events are as follows:\\
\begin{minipage}{.532\textwidth}
	\begin{tcolorbox}[title=Basket, left=4pt, right=1pt]
		\begin{itemize}[leftmargin=*]
			\item $\mathtt{basket}\_\mathtt{activation}(\mathtt{bs})$
			\item $\mathtt{basket}\_\mathtt{addItem}\_\mathtt{execution}(\mathtt{bs})$		
			\item $\mathtt{basket}\_\mathtt{addItem}\_\mathtt{execution}(\mathtt{bs}, \mathtt{name}, \mathtt{price})$		
			\item $\mathtt{basket}\_\mathtt{setName}\_\mathtt{call}(\mathtt{bs})$				
			\item $\mathtt{basket}\_\mathtt{setName}\_\mathtt{call}(\mathtt{bs}, \mathtt{name})$				
			\item $\mathtt{basket}\_\mathtt{setPrice}\_\mathtt{call}(\mathtt{bs})$				
			\item $\mathtt{basket}\_\mathtt{setPrice}\_\mathtt{call}(\mathtt{bs}, \mathtt{price})$				
		\end{itemize}
	\end{tcolorbox}
\end{minipage}
\begin{minipage}{.46\textwidth}
	\begin{tcolorbox}[title=Item, left=4pt, right=1pt]
		\begin{itemize}[leftmargin=*]
			\item $\mathtt{item}\_\mathtt{activation}(\mathtt{it})$
			\item $\mathtt{item}\_\mathtt{setName}\_\mathtt{execution}(\mathtt{it})$		
			\item $\mathtt{item}\_\mathtt{setName}\_\mathtt{execution}(\mathtt{it}, \mathtt{name})$		
			\item $\mathtt{item}\_\mathtt{setPrice}\_\mathtt{execution}(\mathtt{it})$		
			\item $\mathtt{item}\_\mathtt{setPrice}\_\mathtt{execution}(\mathtt{it}, \mathtt{price})$		
			\item $\mathtt{item}\_\mathtt{call}\_\mathtt{basket}\_\mathtt{setName}(\mathtt{it}, \mathtt{bs})$		
			\item $\mathtt{item}\_\mathtt{call}\_\mathtt{basket}\_\mathtt{setPrice}(\mathtt{it}, \mathtt{bs})$
		\end{itemize}
	\end{tcolorbox}
\end{minipage}

\subsection{Generating LTL-formul\ae{} over Events}
Next, we can create LTL-formul\ae{} over the events created in the last step, from the specification of the architectural constraints.
Again, Alg.~\ref{alg:mon} describes a systematic way to do so:
The algorithm essentially modifies architectural assertions by replacing atomic \textsc{FACTum} assertions with corresponding events created by Alg.~\ref{alg:inst}:
To this end, port activations are mapped to corresponding execution/call events without parameters (Lines 5-10),
port valuations to corresponding execution/call events with parameters (Lines 11-16),
component activations to corresponding activation events (Lines 17-18),
and port connections to corresponding call events with source and target locations (Lines 19-20), respectively.

\begin{algorithm}[t]
	\caption{Mapping \textsc{FACTum} to LTL-formul\ae{}}\label{alg:mon}
	\begin{algorithmic}[1]
		\REQUIRE a set $\mathit{AS}$ of architectural constraints
		\ENSURE a set of LTL-formul\ae{}
		\FORALL {$\mathit{\varphi}\in \mathit{AS}$}
		\FORALL {basic assertions $\psi$ in $\varphi$}
		\STATE\COMMENT{assuming $c$ is of type $\mathit{ct}$ and $c'$ is of type $\mathit{ct'}$}
		\SWITCH{$\psi$}
		\CASE{$\caVal{c}{p}$}
		\IF {$p$ is an input port}
		\STATE replace $\psi$ in $\varphi$ with corresponding \emph{execution} event $\mathit{ct}\_p\_\mathtt{execution}(c)$
		\ELSIF{$p$ is an output port}
		\STATE replace $\psi$ in $\varphi$ with corresponding \emph{call} event $\mathit{ct}\_p\_\mathtt{call}(c)$
		\ENDIF
		\ENDCASE
		\CASE{$\caPVal{c}{p}=M$}
		\IF {$p$ is an input port}
		\STATE replace $\psi$ in $\varphi$ with corresponding \emph{execution} event $\mathit{ct}\_p\_\mathtt{execution}(c, M)$
		\ELSIF{$p$ is an output port}
		\STATE replace $\psi$ in $\varphi$ with corresponding \emph{call} event $\mathit{ct}\_p\_\mathtt{call}(c, M)$
		\ENDIF
		\ENDCASE
		\CASE {$\caActive{c}$}
		\STATE replace $\psi$ in $\varphi$ with corresponding \emph{activation} event $\mathit{ct}\_\mathtt{activation}(c)$
		\ENDCASE
		\CASE{$\caConn{c'}{p}{c}{p}$}
		\STATE replace $\psi$ in $\varphi$ with corresponding \emph{call} event $\mathit{ct}\_\mathtt{call}\_\mathit{ct'}\_p(c, c')$
		\ENDCASE
		\ENDSWITCH
		\ENDFOR
		\ENDFOR		
	\end{algorithmic}
\end{algorithm}

From the architectural assertion of our web-shop example (Fig.~\ref{fig:bc:act}), for example, we would generate the following LTL-formula:
\begin{tcolorbox}[ams align*,top=5pt,bottom=4pt]
	&\ctglobally\Big(\mathtt{basket}\_\mathtt{addItem}\_\mathtt{execution}(\mathtt{bs}, \mathtt{n}, \mathtt{p})\Longrightarrow\\
	&\quad\ctnext\big(\mathtt{item}\_\mathtt{activation}(\mathtt{it})\big)\\
	&\quad\land~\ctnext\ctnext\big(\mathtt{basket}\_\mathtt{setPrice}\_\mathtt{call}(\mathtt{bs}, \mathtt{p}) \land \mathtt{basket}\_\mathtt{call}\_\mathtt{item}\_\mathtt{setPrice}(\mathtt{it}, \mathtt{bs})\big)\Big)\\
	&\quad\land~\ctnext\ctnext\ctnext\big(\mathtt{basket}\_\mathtt{setName}\_\mathtt{call}(\mathtt{bs}, \mathtt{n}) \land \mathtt{basket}\_\mathtt{call}\_\mathtt{item}\_\mathtt{setName}(\mathtt{it}, \mathtt{bs})\big)	
\end{tcolorbox}
Note that the atomic \textsc{FACTum} assertions have been replaced by corresponding events described in the last section.

\subsection{Generating Monitors and Code Instrumentation}
The events and LTL-formul\ae{} can finally be used to generate monitors and code instrumentation.
As discussed in Sect.~\ref{sec:rv}, there exist several approaches to automatize this step for different target platforms.

\begin{figure}
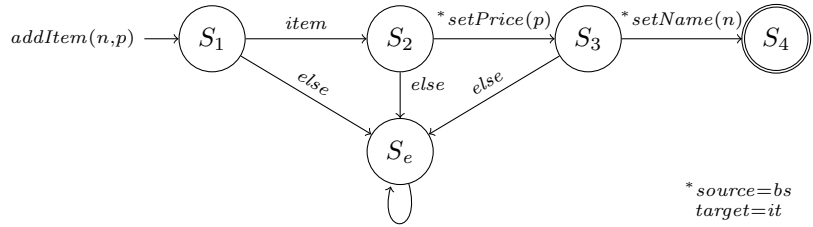

	\centering
	\includestandalone{img/monitor}
	\caption{Monitor for Web shop.\label{fig:monitor}}
\end{figure}

Figure~\ref{fig:monitor} depicts a possible monitor for our web-shop system.
The generated monitor is parameterized by a basket $\mathit{bs}$, an item $\mathit{it}$, name $n$, and price $p$, and starts in state $\mathit{S_1}$, whenever an $\mathit{addItem}$ event is observed with name $n$ and price $p$.
If the next observed event is the creation of a new item, it progresses to state $S_2$; otherwise it moves to an error state $S_e$, in which it remains forever.
From state $\mathit{S_2}$ it either moves to state $S_3$ (for the case the next event is $\mathit{setPrice}$ with price $p$) or in the error state (if it observes any other event).
From state $\mathit{S_3}$ it may again either move to $S_4$ (for the case the next event is $\mathit{setName}$ with name $n$) or the error state.
State $S_4$, however, is a final state, which means that the monitor terminates.
Note that the monitor has no state which signals the successful satisfaction of the property described by Fig.~\ref{fig:bc:act}.
This is because the satisfaction of formula Fig.~\ref{fig:bc:act} can only be determined when observing an infinite trace and never for any finite prefix.

\subsection{Performing the Verification}
After installing the instrumentation code, we can start the monitor to detect architectural violations.

Table~\ref{tab:monitor} depicts a possible run of the web-shop example system (as described in Sect.~\ref{sec:example}) in terms of method calls.
It also lists the corresponding events as received by the monitor (described in Fig.~\ref{fig:monitor}) and the state of the monitor after receiving the event:
\begin{table}
	\centering
	\begin{tabular}{p{4cm}p{4cm}p{2cm}}
		\toprule
		&\multicolumn{2}{c}{Monitor} \\ \cmidrule(r){2-3}
		System Event&Event&State\\
		\midrule	
		\textit{bs}.\textit{addItem(book, 100)} &$\mathit{addItem}(\mathit{book},100)$ & $S_1$ \\
		\textit{it} = \textit{new Item()} &$\mathit{item}$& $S_2$ \\
		\textit{bs} $\longrightarrow$ \textit{it}.\textit{setName(book)} &$\mathit{setName}(\mathit{book})$ & $S_e$ \\
		\textit{bs} $\longrightarrow$ \textit{it}.\textit{setPrice(100)} &$\mathit{setPrice}(100)$ & $S_e$ \\
		\dots & \dots & $S_e$ \\ 
		\bottomrule
	\end{tabular}
	\caption{Possible execution and corresponding state of the monitor.}\label{tab:monitor}
\end{table}
The occurrence of an event $\mathit{addItem}(\mathit{book},\mathit{100})$ triggers the creation of a new monitor which is parameterized with name $\mathit{book}$ and price $100$, and which starts in state $\mathit{S_1}$.
Since the next observed event is the creation of a new item, the monitor moves on to state $\mathit{S_2}$.
In state $S_2$, the monitor expects a $\mathit{setPrice}$ event, however, it observes now a $\mathit{setName}$ event and thus, moves to the error state $\mathit{S_e}$ in which it remains to signal violation of the architectural constraint imposed by the architectural assertion described in Fig.~\ref{fig:bc:act}.
Indeed, if we look at the specification of the assertion, we can see that we required first the price to be set and then the name.
However, if we look at the example code of the \textit{Basket} class, we can see that first the name is initialized and then the price.
Thus, there was indeed a mismatch between the actual specification of the architecture and its implementation which we discovered.

%% file: evaluation.tex
\section{Evaluation}\label{sec:evaluation}
To evaluate the approach, we first implemented it in \textsc{FACTum} Studio~\cite{Marmsoler2018f}.
To this end, we implemented Alg.~\ref{alg:inst} and Alg.~\ref{alg:mon} in XTend~\cite{bettini2016} to generate events and corresponding LTL-formul\ae{} out of a \textsc{FACTum} specification\footnote{The plugin is now part of \textsc{FACTum} Studio and can be downloaded from the corresponding website: \url{https://github.com/habtom/factum/tree/runtimeverification}}.
Then, we applied it to two case studies. The first case study, described in Section \ref{subsec:JetUML} was done in the area of Business Application Systems. The second case study, described in Section \ref{subsec:EmbSys} is performed on Embedded Systems in the automotive domain. 

\subsection{Case Study: Business Application Systems} \label{subsec:JetUML}

\begin{figure}
	\centering
	\includegraphics[width=.55\textwidth]{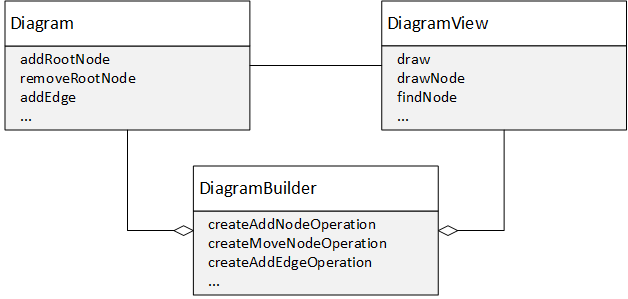}
	\caption{Excerpt of JetUML architecture for the drawing of graphical elements.}\label{fig:jetuml}
\end{figure}

\textit{Study context.}
As a study object in the first case study in the field of Business Application Systems, we chose JetUML~\cite{Robillard2019}, an \emph{open-source} Java application to model UML diagrams.
Its main features consist of creating new diagrams and adding graphical elements to it.
The system's implementation consists of about $35,653$ lines of code, split into $242$ classes.
Figure~\ref{fig:jetuml} shows an excerpt of the architecture of JetUML concerned with the drawing of elements.

\begin{figure}
	\includegraphics[width=\textwidth]{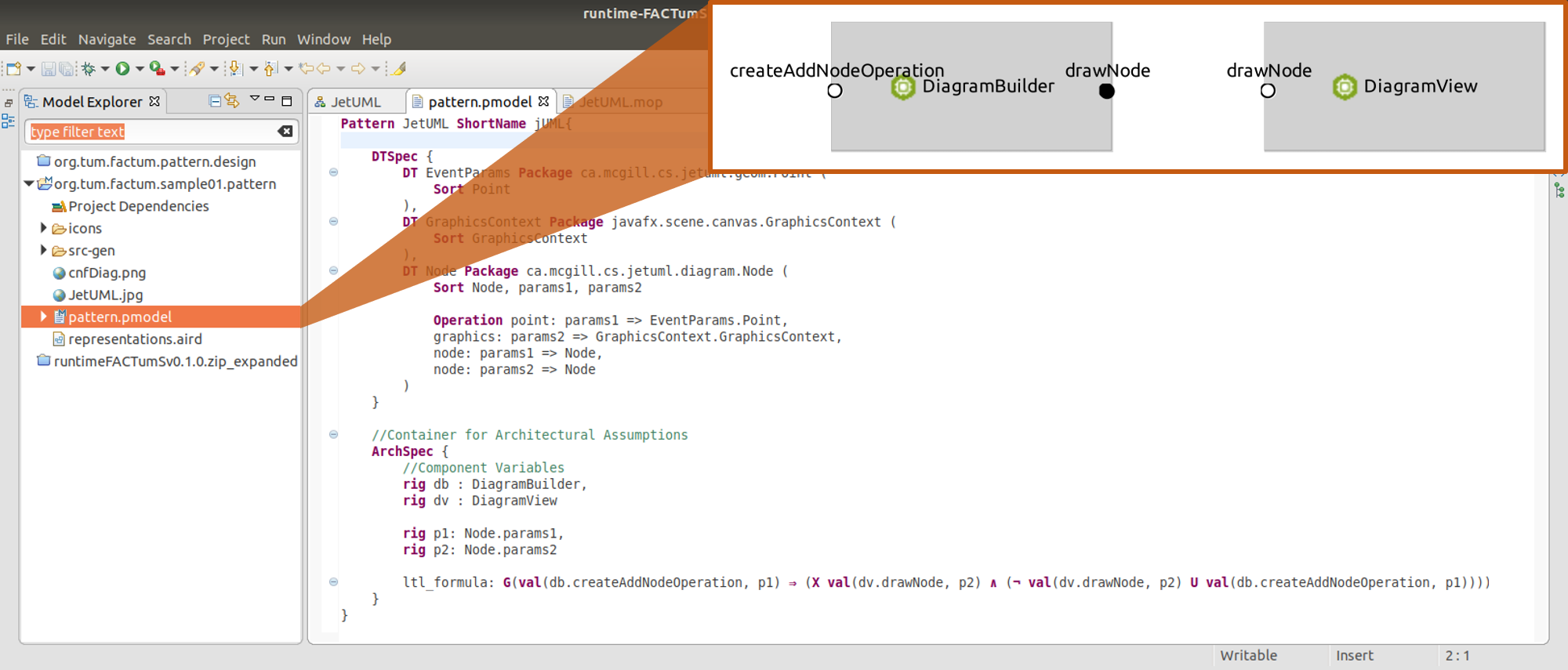}
	\caption{Modeling Architectural Constraints for JetUML in \textsc{FACTum} Studio}\label{fig:juml:factum}
\end{figure}

\textit{Study execution.}
We executed the study in a controlled setting.
We where interested in the parts of the architecture related with the drawing of elements and formulated $10$ different properties of the form:
``Whenever a user adds a new element to the drawing board, a corresponding object is created and drawn by executing the specific method''.
To this end, we formalized the properties in \textsc{FACTum} Studio as shown in Fig.~\ref{fig:juml:factum}.
We then generated events and LTL-formul\ae{} and we used JavaMOP to create corresponding monitors and AspectJ code instrumentations.
Finally, we installed the instrumentation, started the monitors, and observed them for violations of the architecture constraints.

\textit{Findings.}
During the experiments, we could not find any violations of the $10$ assertions specified for this case study.
Thus, we then strengthened the properties to require elements to be drawn only once.
While executing this experiment, the monitors signaled violations of properties.
Namely, after creating the corresponding objects, they were drawn multiple times.
While this is not a severe bug, it can indeed be considered as a design issue, since it involves unnecessary computation which might decrease performance.

\subsection{Case Study: Embedded Systems} \label{subsec:EmbSys}

\begin{figure}
	\centering
	\includegraphics[width=8cm,height=5cm]{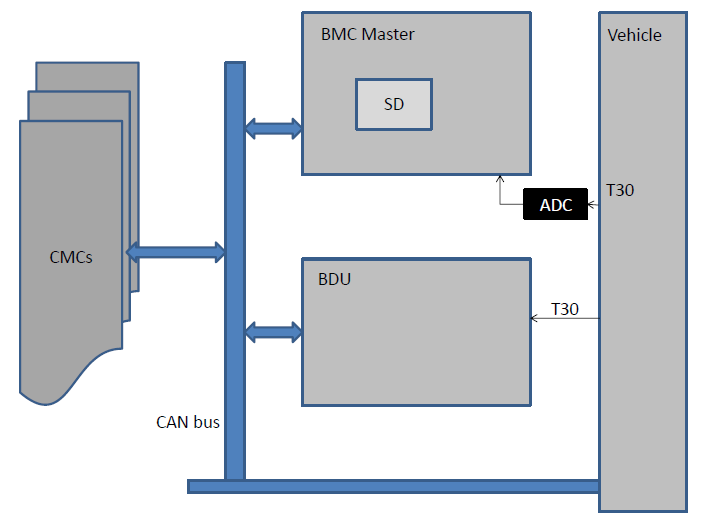}
	\caption{Architecture of the Battery Management system}\label{fig:case2:arch}
\end{figure}

\textit{Study context.}
The second case study was executed in collaboration with an industrial partner from the automotive domain.
In this use case we analyzed the architecture of the Service Disconnect (SD) software component in the Battery Management System (BMS) in the vehicles. The system is responsible for monitoring and regulating a car's battery.
Figure~\ref{fig:case2:arch} depicts the simplified form of the BMC architecture provided by the partner and contains the following components:\\
Component \emph{Vehicle} is an abstraction of the car itself.\\
\emph{Analog to digital converter (ADC)} forwards the signal to component \emph{BMC Master}. The ADC pin can additionally be used for checking the service disconnect status. \\
\emph{Battery disconnect unit (BDU)} monitors the connection of the car to its battery. It communicates the status to \emph{BMC Master} via the CAN bus. \\
\emph{BMC Master} is the actual battery management component which provides a functionality ``Service Disconnect'' which communicates the connection state of the battery to the vehicle.
To this end, it compares the values obtained from \emph{ADC} and \emph{BDU}. For the case that the values are the same, it forwards it to the car. If the two signals differ from each other, then \emph{BMC Master} should send an error message. In \emph{BMC Master} reside all battery management related
software features. One of those features is SD whose dynamic properties we verify at runtime. 

\begin{figure}[t]
	\includegraphics[width=\textwidth]{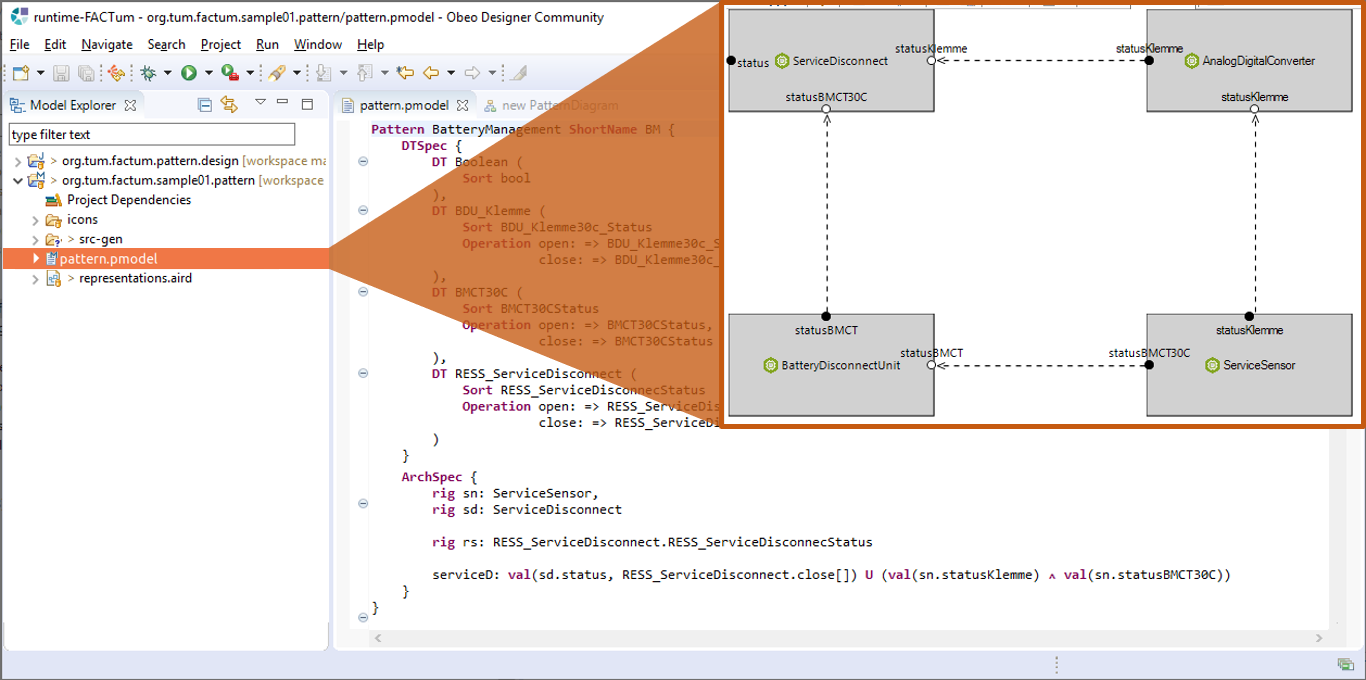}
	\caption{Modeling Architectural Constraints for Battery Management System in \textsc{FACTum} Studio.}\label{fig:battery}
\end{figure}

\textit{Study execution.}
In contrast to the first case study, this study was executed in an industrial setting in collaboration with an industrial partner.
We where interested in analyzing the following constraint for the architecture:
``The Battery is signaled to be disconnected only if BDU signals disconnect and ADC signals disconnect''.
Again, we first formalized the property in \textsc{FACTum} Studio (Fig.~\ref{fig:battery}) and generated events and LTL-formul\ae{}.
This time, however, code instrumentation was done using CAPL script.
The log communication from the components' signals was obtained through CANoe tool.
The monitor was written in C\# and created using LTL3 tools~\cite{Bauer2019}.
Finally, the system was tested, and the collected log-files were inspected using the previously created monitor.

\textit{Findings.}
In this case study we specified four properties in total and our experiments revealed runs of the system in which the architecture property was indeed violated.
After communicating our results to our industry partner, they confirmed that the architecture specification was wrong.
The reason was indeed a so-called architectural erosion:
Over time, the system was adapted and communication over the ADC connector was replaced with a direct communication over the CAN bus.
However, this is not reflected in the architecture, which still shows the original connection through the ADC connector.

%% file: rw.tex
\section{Related Work}\label{sec:rw}

In this paper, we provide a systematic approach that detects architectural erosion or architectural drift based on runtime verification\footnote{In this paper the difference between architectural erosion and architectural drift is not considered.}. Although different techniques for controlling software architectural erosion have been proposed across the literature, previous work has mainly focused on static analysis methods. In this paper, instead of applying static analysis, we present a new approach for solving the architectural erosion problem by applying runtime verification. This allows us to go beyond detecting static violations of the systems, but rather focusing and checking dynamic violations that are emerging from the architectural dynamicity of the systems. Additionally, dynamic software analysis approaches or runtime verification is vastly used to solve various problems in many different fields. However, until now it has not been applied to ensure architecture consistency. We believe that our approach is the first one which combines and utilizes RV for the analysis of dynamic architectural drift. Therefore, in this section, we discuss related work on the field of architectural erosion and runtime verification.

Murphy at al.~\cite{murphy1995software}, Koschke and Simon~\cite{koschke2003hierarchical} and Said et al.~
\cite{said2018reflexion} propose reflection model techniques as architectural solutions for controlling the software erosion. The reflection model techniques compares a model of the implemented architecture and a hypothetical model of the intended architecture. The latter is created from a static analysis of the source code. 

Lavery and Watanabe in~\cite{Lavery2017} present a runtime monitoring method for actor-based programs and a scala-based asynchronous runtime-monitoring module that realizes the proposed method. They aim to provide failure recovery and mitigation mechanism for Scala applications by making use of the lightweight software to monitor the properties specified. The module does not require specialized languages for describing application properties that need to be monitored. The programmer specifies in Scala the property that needs to be verified, and the mitigation code that needs to be invoked when a particular property is violated.

To make dynamic reconfigurations more reliable, L{\'e}ger et al.~\cite{leger2010reliable} proposes an approach ensuring that system consistency and availability is maintained despite run-time failures and changes in the system. A reconfiguration is a modification of a system state during its execution, and it may potentially put this system in an inconsistent state.  In the first step the authors provide a model of configurations and reconfigurations. They specify consistency by means of integrity constraints, i.e. configuration invariants and pre/post-conditions on reconfiguration operations. Alloy has been used as a specification language to model these constraints, which are later translated in FPath, a navigation language used as a constraint language in Fractal architectures to check the validity of integrity constraints on real systems at runtime.


%% file: conclusion.tex
\section{Conclusion}\label{sec:conc}
In this paper, we present an approach that provides a solution to the problem of detecting architectural erosion for dynamic architectures.
To this end, architectural constraints are formally specified using \textsc{FACTum}, a language for the specification of constraints for dynamic architectures, and then systematically transferred to corresponding monitors and code instrumentation which can be used to detect violations at runtime.

In the paper, we describe the approach and demonstrate its applicability through a running example.
Next, we describe two algorithms which can be used to generate code instrumentations and monitors, which monitor the architectural violations from the \textsc{FACTum} specification.
Finally, we describe the outcome of two case studies on which we evaluated the proposed approach in the context of an open source Java application and a proprietary C application.

Our results suggest that runtime verification is indeed feasible to detect architectural erosion for different types of applications: from embedded C applications to object-oriented business information systems.
Additionally, our evaluations of the approach that we propose in this paper show that it scales well and has the potential to uncover important architecture violations.

However, our results also expose some limitations of the approach.
First, our approach can only be used to detect violations and not to guarantee the absence of architectural violations, nor to act whenever an incorrect behavior is detected.
Second, it is not yet possible to analyze real-time requirements. This posed a serious limitation, particularly for the second use case,  since many important architectural assertions require timed aspects.\looseness-1

The first limitation is a general limitation of runtime verification, and there is not much we can do about this.
However, the second limitation can be addressed in the future, as future work.
To this end, future work should focus to extend the approach with timing aspects.\looseness-1

\textit{Acknowledgements.}
We want to thank Ilias Gerostathopoulos for helpful discussions on architectural erosion.
Parts of the work on which we report in this paper was funded by the German Federal Ministry of Education and Research (BMBF) under grant no. 01Is16043A.

%% file: main.bbl
\begin{thebibliography}{10}
\providecommand{\bibitemdeclare}[2]{}
\providecommand{\surnamestart}{}
\providecommand{\surnameend}{}
\providecommand{\urlprefix}{Available at }
\providecommand{\url}[1]{\texttt{#1}}
\providecommand{\href}[2]{\texttt{#2}}
\providecommand{\urlalt}[2]{\href{#1}{#2}}
\providecommand{\doi}[1]{doi:\urlalt{http://dx.doi.org/#1}{#1}}
\providecommand{\bibinfo}[2]{#2}

\bibitemdeclare{}{Bauer2019}
\bibitem{Bauer2019}
\bibinfo{author}{Andreas \surnamestart Bauer\surnameend}
  (\bibinfo{year}{2019}): \emph{\bibinfo{title}{LTL3 Tools}}.
\newblock \urlprefix\url{http://ltl3tools.sourceforge.net/}.

\bibitemdeclare{inproceedings}{bauer2006model}
\bibitem{bauer2006model}
\bibinfo{author}{Andreas \surnamestart Bauer\surnameend},
  \bibinfo{author}{Martin \surnamestart Leucker\surnameend} \&
  \bibinfo{author}{Christian \surnamestart Schallhart\surnameend}
  (\bibinfo{year}{2006}): \emph{\bibinfo{title}{Model-based runtime analysis of
  distributed reactive systems}}.
\newblock In: {\sl \bibinfo{booktitle}{Software Engineering Conference, 2006.
  Australian}}, \bibinfo{organization}{IEEE}, pp. \bibinfo{pages}{10--pp},
  \doi{10.1109/aswec.2006.36}.

\bibitemdeclare{inproceedings}{bauer2006monitoring}
\bibitem{bauer2006monitoring}
\bibinfo{author}{Andreas \surnamestart Bauer\surnameend},
  \bibinfo{author}{Martin \surnamestart Leucker\surnameend} \&
  \bibinfo{author}{Christian \surnamestart Schallhart\surnameend}
  (\bibinfo{year}{2006}): \emph{\bibinfo{title}{Monitoring of real-time
  properties}}.
\newblock In: {\sl \bibinfo{booktitle}{International Conference on Foundations
  of Software Technology and Theoretical Computer Science}},
  \bibinfo{organization}{Springer}, pp. \bibinfo{pages}{260--272},
  \doi{10.1007/11944836_25}.

\bibitemdeclare{article}{bauer2011runtime}
\bibitem{bauer2011runtime}
\bibinfo{author}{Andreas \surnamestart Bauer\surnameend},
  \bibinfo{author}{Martin \surnamestart Leucker\surnameend} \&
  \bibinfo{author}{Christian \surnamestart Schallhart\surnameend}
  (\bibinfo{year}{2011}): \emph{\bibinfo{title}{Runtime verification for LTL
  and TLTL}}.
\newblock {\sl \bibinfo{journal}{ACM Transactions on Software Engineering and
  Methodology (TOSEM)}} \bibinfo{volume}{20}(\bibinfo{number}{4}),
  p.~\bibinfo{pages}{14}, \doi{10.1145/2000799.2000800}.

\bibitemdeclare{book}{bettini2016}
\bibitem{bettini2016}
\bibinfo{author}{Lorenzo \surnamestart Bettini\surnameend}
  (\bibinfo{year}{2016}): \emph{\bibinfo{title}{Implementing domain-specific
  languages with Xtext and Xtend}}.
\newblock \bibinfo{publisher}{Packt Publishing Ltd}.

\bibitemdeclare{inproceedings}{Broy1996}
\bibitem{Broy1996}
\bibinfo{author}{Manfred \surnamestart Broy\surnameend} (\bibinfo{year}{1996}):
  \emph{\bibinfo{title}{Algebraic Specification of Reactive Systems}}.
\newblock In: {\sl \bibinfo{booktitle}{Algebraic Methodology and Software
  Technology}}, \bibinfo{organization}{Springer}, \bibinfo{publisher}{Springer
  Berlin Heidelberg}, pp. \bibinfo{pages}{487--503}, \doi{10.1007/bfb0014335}.

\bibitemdeclare{inproceedings}{chen2005java}
\bibitem{chen2005java}
\bibinfo{author}{Feng \surnamestart Chen\surnameend} \&
  \bibinfo{author}{Grigore \surnamestart Ro{\c{s}}u\surnameend}
  (\bibinfo{year}{2005}): \emph{\bibinfo{title}{Java-MOP: A monitoring oriented
  programming environment for Java}}.
\newblock In: {\sl \bibinfo{booktitle}{International Conference on Tools and
  Algorithms for the Construction and Analysis of Systems}},
  \bibinfo{organization}{Springer}, pp. \bibinfo{pages}{546--550},
  \doi{10.1007/978-3-540-31980-1_36}.

\bibitemdeclare{inproceedings}{chen2007mop}
\bibitem{chen2007mop}
\bibinfo{author}{Feng \surnamestart Chen\surnameend} \&
  \bibinfo{author}{Grigore \surnamestart Ro{\c{s}}u\surnameend}
  (\bibinfo{year}{2007}): \emph{\bibinfo{title}{Mop: an efficient and generic
  runtime verification framework}}.
\newblock In: {\sl \bibinfo{booktitle}{Acm Sigplan Notices}},
  \bibinfo{volume}{42}, \bibinfo{organization}{ACM}, pp.
  \bibinfo{pages}{569--588}, \doi{10.1145/1297105.1297069}.

\bibitemdeclare{book}{clarke1999model}
\bibitem{clarke1999model}
\bibinfo{author}{EM~\surnamestart Clarke\surnameend}, \bibinfo{author}{Orna
  \surnamestart Grumberg\surnameend} \& \bibinfo{author}{Doron~A \surnamestart
  Peled\surnameend} (\bibinfo{year}{1999}): \emph{\bibinfo{title}{Model
  Checking}}.
\newblock \bibinfo{publisher}{Cambridge, Mass, MIT Press}.

\bibitemdeclare{}{Coverity2019}
\bibitem{Coverity2019}
\bibinfo{author}{\surnamestart Coverity\surnameend} (\bibinfo{year}{2019}):
  \emph{\bibinfo{title}{Architecture analysis}}.
\newblock
  \urlprefix\url{http://www.coverity.com/products/architectureanalysis}.

\bibitemdeclare{inproceedings}{Deissenboeck2010}
\bibitem{Deissenboeck2010}
\bibinfo{author}{Florian \surnamestart Deissenboeck\surnameend},
  \bibinfo{author}{Lars \surnamestart Heinemann\surnameend},
  \bibinfo{author}{Benjamin \surnamestart Hummel\surnameend} \&
  \bibinfo{author}{Elmar \surnamestart J{\"{u}}rgens\surnameend}
  (\bibinfo{year}{2010}): \emph{\bibinfo{title}{Flexible architecture
  conformance assessment with ConQAT}}.
\newblock In: {\sl \bibinfo{booktitle}{Proceedings of the 32nd {ACM/IEEE}
  International Conference on Software Engineering - Volume 2, {ICSE} 2010,
  Cape Town, South Africa, 1-8 May 2010}}, pp. \bibinfo{pages}{247--250},
  \doi{10.1145/1810295.1810343}.

\bibitemdeclare{article}{Eick2001}
\bibitem{Eick2001}
\bibinfo{author}{S.~G. \surnamestart Eick\surnameend}, \bibinfo{author}{T.~L.
  \surnamestart Graves\surnameend}, \bibinfo{author}{A.~F. \surnamestart
  Karr\surnameend}, \bibinfo{author}{J.~S. \surnamestart Marron\surnameend} \&
  \bibinfo{author}{A.~\surnamestart Mockus\surnameend} (\bibinfo{year}{2001}):
  \emph{\bibinfo{title}{Does code decay? Assessing the evidence from change
  management data}}.
\newblock {\sl \bibinfo{journal}{IEEE Transactions on Software Engineering}}
  \bibinfo{volume}{27}(\bibinfo{number}{1}), pp. \bibinfo{pages}{1--12},
  \doi{10.1109/32.895984}.

\bibitemdeclare{incollection}{emerson2008beginning}
\bibitem{emerson2008beginning}
\bibinfo{author}{E~Allen \surnamestart Emerson\surnameend}
  (\bibinfo{year}{2008}): \emph{\bibinfo{title}{The beginning of model
  checking: A personal perspective}}.
\newblock In: {\sl \bibinfo{booktitle}{25 Years of Model Checking}},
  \bibinfo{publisher}{Springer}, pp. \bibinfo{pages}{27--45},
  \doi{10.1007/978-3-540-69850-0_2}.

\bibitemdeclare{misc}{Foundation2019}
\bibitem{Foundation2019}
\bibinfo{author}{Eclipse \surnamestart Foundation\surnameend}
  (\bibinfo{year}{2019}): \emph{\bibinfo{title}{AspectJ}}.
\newblock \bibinfo{howpublished}{\url{https://www.eclipse.org/aspectj/}}.

\bibitemdeclare{article}{Garlan1995}
\bibitem{Garlan1995}
\bibinfo{author}{David \surnamestart Garlan\surnameend},
  \bibinfo{author}{Robert \surnamestart Allen\surnameend} \&
  \bibinfo{author}{John \surnamestart Ockerbloom\surnameend}
  (\bibinfo{year}{1995}): \emph{\bibinfo{title}{Architectural Mismatch: Why
  Reuse Is So Hard}}.
\newblock {\sl \bibinfo{journal}{IEEE Softw.}}
  \bibinfo{volume}{12}(\bibinfo{number}{6}), pp. \bibinfo{pages}{17--26},
  \doi{10.1109/52.469757}.

\bibitemdeclare{inproceedings}{Godfrey2000}
\bibitem{Godfrey2000}
\bibinfo{author}{Michael~W. \surnamestart Godfrey\surnameend} \&
  \bibinfo{author}{Eric H.~S. \surnamestart Lee\surnameend}
  (\bibinfo{year}{2000}): \emph{\bibinfo{title}{Secrets from the Monster:
  Extracting Mozilla's Software Architecture}}.
\newblock In: {\sl \bibinfo{booktitle}{In Proc. of 2000 Intl. Symposium on
  Constructing software engineering tools (CoSET 2000}}, pp.
  \bibinfo{pages}{15--23}.

\bibitemdeclare{}{Klocwork2019}
\bibitem{Klocwork2019}
\bibinfo{author}{\surnamestart Klocwork\surnameend} (\bibinfo{year}{2019}):
  \emph{\bibinfo{title}{Klockwork Architect}}.
\newblock
  \urlprefix\url{http://www.klocwork.com/products/insight/architect-code-visualization/}.

\bibitemdeclare{inproceedings}{koschke2003hierarchical}
\bibitem{koschke2003hierarchical}
\bibinfo{author}{R.~\surnamestart {Koschke}\surnameend} \&
  \bibinfo{author}{D.~\surnamestart {Simon}\surnameend} (\bibinfo{year}{2003}):
  \emph{\bibinfo{title}{Hierarchical reflexion models}}.
\newblock In: {\sl \bibinfo{booktitle}{10th Working Conference on Reverse
  Engineering, 2003. WCRE 2003. Proceedings.}}, pp. \bibinfo{pages}{36--45},
  \doi{10.1109/WCRE.2003.1287235}.

\bibitemdeclare{}{Lattix2019}
\bibitem{Lattix2019}
\bibinfo{author}{\surnamestart Lattix\surnameend} (\bibinfo{year}{2019}):
  \emph{\bibinfo{title}{The Lattix architecture management system}}.
\newblock \urlprefix\url{http://www.lattix.com/products}.

\bibitemdeclare{inproceedings}{Lavery2017}
\bibitem{Lavery2017}
\bibinfo{author}{P.~\surnamestart Lavery\surnameend} \&
  \bibinfo{author}{T.~\surnamestart Watanabe\surnameend}
  (\bibinfo{year}{2017}): \emph{\bibinfo{title}{An actor-based runtime
  monitoring system for web and desktop applications}}.
\newblock In: {\sl \bibinfo{booktitle}{2017 18th IEEE/ACIS International
  Conference on Software Engineering, Artificial Intelligence, Networking and
  Parallel/Distributed Computing (SNPD)}}, pp. \bibinfo{pages}{385--390},
  \doi{10.1109/SNPD.2017.8022750}.

\bibitemdeclare{inproceedings}{leger2010reliable}
\bibitem{leger2010reliable}
\bibinfo{author}{Marc \surnamestart L{\'e}ger\surnameend},
  \bibinfo{author}{Thomas \surnamestart Ledoux\surnameend} \&
  \bibinfo{author}{Thierry \surnamestart Coupaye\surnameend}
  (\bibinfo{year}{2010}): \emph{\bibinfo{title}{Reliable dynamic
  reconfigurations in a reflective component model}}.
\newblock In: {\sl \bibinfo{booktitle}{International Symposium on
  Component-Based Software Engineering}}, \bibinfo{organization}{Springer}, pp.
  \bibinfo{pages}{74--92}, \doi{10.1007/978-3-642-13238-4_5}.

\bibitemdeclare{article}{leucker2009brief}
\bibitem{leucker2009brief}
\bibinfo{author}{Martin \surnamestart Leucker\surnameend} \&
  \bibinfo{author}{Christian \surnamestart Schallhart\surnameend}
  (\bibinfo{year}{2009}): \emph{\bibinfo{title}{A brief account of runtime
  verification}}.
\newblock {\sl \bibinfo{journal}{The Journal of Logic and Algebraic
  Programming}} \bibinfo{volume}{78}(\bibinfo{number}{5}), pp.
  \bibinfo{pages}{293--303}, \doi{10.1016/j.jlap.2008.08.004}.

\bibitemdeclare{}{structure1012019}
\bibitem{structure1012019}
\bibinfo{author}{Headway Software~Technologies \surnamestart Ltd.\surnameend}
  (\bibinfo{year}{2019}): \emph{\bibinfo{title}{structure101}}.
\newblock \urlprefix\url{https://structure101.com/}.

\bibitemdeclare{article}{Marmsoler2016a}
\bibitem{Marmsoler2016a}
\bibinfo{author}{D.~\surnamestart Marmsoler\surnameend} \&
  \bibinfo{author}{M.~\surnamestart Gleirscher\surnameend}
  (\bibinfo{year}{2016}): \emph{\bibinfo{title}{On Activation, Connection, and
  Behavior in Dynamic Architectures}}.
\newblock {\sl \bibinfo{journal}{Scientific Annals of Computer Science}}
  \bibinfo{volume}{26}(\bibinfo{number}{2}), p. \bibinfo{pages}{187–248},
  \doi{10.7561/SACS.2016.2.187}.

\bibitemdeclare{inproceedings}{Marmsoler2018c}
\bibitem{Marmsoler2018c}
\bibinfo{author}{Diego \surnamestart Marmsoler\surnameend}
  (\bibinfo{year}{2018}): \emph{\bibinfo{title}{Hierarchical Specication and
  Verication of Architecture Design Patterns}}.
\newblock In: {\sl \bibinfo{booktitle}{Fundamental Approaches to Software
  Engineering - 21th International Conference, {FASE} 2018, Held as Part of the
  European Joint Conferences on Theory and Practice of Software, {ETAPS} 2018,
  Thessaloniki, Greece, April 14-20, 2018, Proceedings}},
  \doi{10.1007/978-3-319-89363-1_9}.

\bibitemdeclare{phdthesis}{Marmsoler2019}
\bibitem{Marmsoler2019}
\bibinfo{author}{Diego \surnamestart Marmsoler\surnameend}
  (\bibinfo{year}{2019}): \emph{\bibinfo{title}{Axiomatic Specification and
  Interactive Verification of Architectural Design Patterns in FACTum}}.
\newblock \bibinfo{type}{Dissertation}, \bibinfo{school}{Technische
  Universität München}, \bibinfo{address}{München}.

\bibitemdeclare{inproceedings}{Marmsoler2018f}
\bibitem{Marmsoler2018f}
\bibinfo{author}{Diego \surnamestart Marmsoler\surnameend} \&
  \bibinfo{author}{Habtom~Kahsay \surnamestart Gidey\surnameend}
  (\bibinfo{year}{2018}): \emph{\bibinfo{title}{\textsc{FACTum} {S}tudio: A
  Tool for the Axiomatic Specification and Verification of Architectural Design
  Patterns}}.
\newblock In: {\sl \bibinfo{booktitle}{Formal Aspects of Component Software -
  {FACS} 2018 - 15th International Conference, Proceedings}},
  \doi{10.1007/978-3-030-02146-7_14}.

\bibitemdeclare{article}{Marmsoler2019a}
\bibitem{Marmsoler2019a}
\bibinfo{author}{Diego \surnamestart Marmsoler\surnameend} \&
  \bibinfo{author}{Habtom~Kashay \surnamestart Gidey\surnameend}
  (\bibinfo{year}{2019}): \emph{\bibinfo{title}{Interactive verification of
  architectural design patterns in FACTum}}.
\newblock {\sl \bibinfo{journal}{Formal Aspects of Computing}},
  \doi{10.1007/s00165-019-00488-x}.

\bibitemdeclare{incollection}{Marmsoler2016}
\bibitem{Marmsoler2016}
\bibinfo{author}{Diego \surnamestart Marmsoler\surnameend} \&
  \bibinfo{author}{Mario \surnamestart Gleirscher\surnameend}
  (\bibinfo{year}{2016}): \emph{\bibinfo{title}{Specifying Properties of
  Dynamic Architectures using Configuration Traces}}.
\newblock In: {\sl \bibinfo{booktitle}{International Colloquium on Theoretical
  Aspects of Computing}}, \bibinfo{publisher}{Springer}, pp.
  \bibinfo{pages}{235--254}, \doi{10.1007/978-3-319-46750-4_14}.

\bibitemdeclare{article}{murphy1995software}
\bibitem{murphy1995software}
\bibinfo{author}{Gail~C \surnamestart Murphy\surnameend},
  \bibinfo{author}{David \surnamestart Notkin\surnameend} \&
  \bibinfo{author}{Kevin \surnamestart Sullivan\surnameend}
  (\bibinfo{year}{1995}): \emph{\bibinfo{title}{Software reflexion models:
  Bridging the gap between source and high-level models}}.
\newblock {\sl \bibinfo{journal}{ACM SIGSOFT Software Engineering Notes}}
  \bibinfo{volume}{20}(\bibinfo{number}{4}), pp. \bibinfo{pages}{18--28},
  \doi{10.1145/222132.222136}.

\bibitemdeclare{inproceedings}{pnueli1977temporal}
\bibitem{pnueli1977temporal}
\bibinfo{author}{Amir \surnamestart Pnueli\surnameend} (\bibinfo{year}{1977}):
  \emph{\bibinfo{title}{The temporal logic of programs}}.
\newblock In: {\sl \bibinfo{booktitle}{Foundations of Computer Science, 1977.,
  18th Annual Symposium on}}, \bibinfo{organization}{IEEE}, pp.
  \bibinfo{pages}{46--57}, \doi{10.1109/sfcs.1977.32}.

\bibitemdeclare{}{Robillard2019}
\bibitem{Robillard2019}
\bibinfo{author}{Martin \surnamestart Robillard\surnameend}
  (\bibinfo{year}{2019}): \emph{\bibinfo{title}{JetUML}}.
\newblock \urlprefix\url{https://github.com/prmr/JetUML}.

\bibitemdeclare{}{Runtimeverification2019}
\bibitem{Runtimeverification2019}
\bibinfo{author}{\surnamestart Runtimeverification\surnameend}
  (\bibinfo{year}{2019}): \emph{\bibinfo{title}{JavaMOP}}.
\newblock \urlprefix\url{https://github.com/runtimeverification/javamop}.

\bibitemdeclare{inproceedings}{said2018reflexion}
\bibitem{said2018reflexion}
\bibinfo{author}{Wasim \surnamestart Said\surnameend}, \bibinfo{author}{Jochen
  \surnamestart Quante\surnameend} \& \bibinfo{author}{Rainer \surnamestart
  Koschke\surnameend} (\bibinfo{year}{2018}): \emph{\bibinfo{title}{Reflexion
  Models for State Machine Extraction and Verification}}.
\newblock In: {\sl \bibinfo{booktitle}{2018 IEEE International Conference on
  Software Maintenance and Evolution (ICSME)}}, \bibinfo{organization}{IEEE},
  pp. \bibinfo{pages}{149--159}, \doi{10.1109/icsme.2018.00025}.

\bibitemdeclare{}{Sonargraph2019}
\bibitem{Sonargraph2019}
\bibinfo{author}{\surnamestart Sonargraph\surnameend} (\bibinfo{year}{2019}):
  \emph{\bibinfo{title}{Sonargraph-Architect}}.
\newblock \urlprefix\url{http://www.hello2morrow.com/products/sonargraph}.

\bibitemdeclare{incollection}{Wirsing1990}
\bibitem{Wirsing1990}
\bibinfo{author}{Martin \surnamestart Wirsing\surnameend}
  (\bibinfo{year}{1990}): \emph{\bibinfo{title}{Algebraic Specification}}.
\newblock In \bibinfo{editor}{Jan \surnamestart van Leeuwen\surnameend},
  editor: {\sl \bibinfo{booktitle}{Handbook of Theoretical Computer Science
  (Vol. B)}}, \bibinfo{publisher}{MIT Press}, \bibinfo{address}{Cambridge, MA,
  USA}, pp. \bibinfo{pages}{675--788},
  \doi{10.1016/b978-0-444-88074-1.50018-4}.

\end{thebibliography}
